# Transition from static to dynamic macroscopic friction in the framework of the Frenkel-Kontorova model

Naum I. Gershenzon[1,2], Gust Bambakidis[1]

[1]Physics Department, Wright State University, 3640 Colonel Glenn Highway Dayton, OH 45435

[2]Department of Earth and Environmental Sciences, Wright State University, 3640 Colonel Glenn Highway  Dayton, OH 45435

**Abstract**

A new generation of experiments on dry macroscopic friction has revealed that the transition from static to dynamic friction is essentially a spatially and temporally non-uniform process, initiated by a rupture-like detachment front. We show the suitability of the Frenkel-Kontorova model for describing this transition. The model predicts the existence of two types of detachment fronts, explaining both the variability and abrupt change of velocity observed in experiments. The quantitative relation obtained between the velocity of the detachment front and the ratio of shear to normal stress is consistent with experiments. The model provides a functional dependence between slip velocity and shear stress, and predicts that slip velocity is independent of normal stress. Paradoxically, the transition from static to dynamic friction does not depend explicitly on ether the static or the dynamic friction coefficient, although the beginning and end of transition process are controlled by these coefficients.

1. Introduction

An understanding of tribology and its complicated nonlinear aspects requires a combination of experimental, theoretical and computational efforts [1]. While experiments and molecular-dynamic (MD) simulations provides invaluable information about the atomic origins of static and dynamic friction, the complexities of realistic 3D systems makes it difficult to understand the general mechanisms underlying friction. In this regard, simple low-dimensional phenomenological models, such as the Tomlinson [2] and Frenkel–Kontorova (FK) [3] models are useful tools for determining the essential features of nonlinear sliding phenomena [4]. These features can then be tested by experiment and MD simulation. Thus, development of such models is an essential part of studying friction.

The relation between microscopic and macroscopic friction has yet to be established. But in this article we demonstrate that the FK model, which has been widely used by others to study both micro- and nanoscopic friction [5, 6], may also describe the dynamics of macroscopic dry friction as well. The motivation for this is the similarity between plasticity and dry friction, both on laboratory and geophysical scales. Plastic deformation (the relative slip of two parts of a crystal) occurs due to the movement of edge dislocations by the temporally and spatially localized shift of crystalline planes. In this case the external stress initiating plasticity is only a small fraction of the stress necessary for the uniform relative displacement of planes of crystal atoms. The same is true for spasmodic local motion along faults in the Earth's crust, which occurs during earthquakes, creep and slow slip events. The results of a new generation of laboratory macroscopic friction experiments also make it clear that these frictional processes are essentially <u>non-uniform in time and space</u> [7-10]. In the model we propose, sliding occurs in much the same way as plasticity, i.e. due to <u>movement of a certain type of defect</u> (a "macroscopic dislocation") nucleated on the frictional surfaces by shear stress in the presence of asperities. This is a basic distinction between our approach to macroscopic friction and other block-spring models [11].

Originally, the FK model was introduced for systems with a periodic potential. The spatial distribution of asperities over a frictional surface is usually not periodic. We suppose that the basic features of the FK model, e.g. soliton-like solutions, are still preserved even in the case of a randomized distribution of asperity sizes centred about some characteristic value. This supposition is supported by studies of the FK model and its continuum limit, i.e. the sine-Gordon (SG) equation, with 1) a quasi-periodic potential [5], 2) a potential with impurities [12] and 3) a randomized potential [13].

In a series of experiments with poly(methyl-methacrylate) (PMMA) material it has been shown [7-10] (see Fig. 1a for a schematic of experiments) that:
1. a series of slip pulses (precursors), occurring under shear stress far below the value necessary to overcome static friction, precedes sliding of the PMMA blocks (Fig. 4);
2. the transition from static to dynamic friction occurs due to the appearance of a so-called detachment front (a sudden change of the actual contact area between frictional surfaces) propagating with velocities ranging from a few percent of the Rayleigh wave speed (Sub-Rayleigh and "slow" fronts) to beyond shear wave (super-shear) velocities; the detachment front precedes slip;
3. the magnitude of the front velocity may vary in time and space and may instantaneously be reduced to several sub-multiples of its value;
4. there is a functional dependence of detachment front speed on the ratio of shear to normal stress (see Fig. 5).

The goal of this article is a to describe qualitatively and quantitatively the above transition process. We start with a model (part 2), then consider solutions of the SG equation appropriate for our problem (part 3). Based on the latter we model (part 4) results of some experiments reported previously by others [7-10]. The applicability of the model to the description of the dynamics of earthquakes is illustrated in part 5.

## 2. Model description

To apply the FK model we will consider the asperities (together with the surrounding material) (Fig. 1b) on one of the frictional surfaces as forming a linear chain of balls of mass $M$, each ball interacting with the nearest neighbours on either side via spring forces of constant $K_b$ (Fig. 1c). The asperities on the opposite frictional surface will be regarded as forming a rigid substrate which interacts with the masses $M$ via a periodic potential. Then we can apply the one-dimensional (1D) FK model to describe the slip dynamics:

$$M \frac{\partial^2 u_i}{\partial t^2} - K_b (u_{i+1} - 2u_i + u_{i-1}) + F_d \sin \frac{2\pi}{b} u_i = F(x,t) - f_i(x,t,\frac{\partial u_i}{\partial t}), \qquad (1)$$

where $u_i$ is the shift of ball $i$ relative to its equilibrium position, $b$ is a typical distance between asperities, $t$ is time, $F_d$ is the amplitude of the periodic force on $M$ associated with the periodic substrate potential, $f_i$ is the frictional force on asperity $i$, and $F_i$ is the external force. Of course, a 2D chain is more realistic, but 1D is also informative, in the same way as in the theory of dislocations.

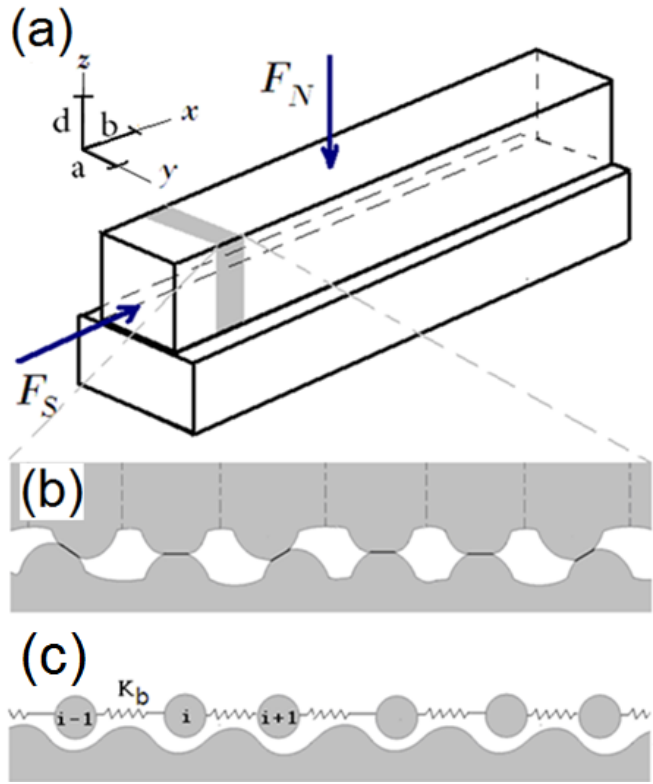

Figure 1. Schematic of a) experimental arrangement [7-10], b) asperity contact and c) chain of masses interacting via elastic springs and placed in a periodic potential (substrate).

In previous work eq. (1) was applied to study plasticity in crystalline materials, which involves the dynamics of atomic-scale edge dislocations [14-16]. To express the coefficients of eq. (1) in terms of the volume and surface mechanical parameters of the frictional blocks and external conditions such as normal stress, we first consider these coefficients for the case of plasticity. So we assume a sliding surface parallel to the actual frictional surface but inside the block. Supposing that it is a crystal material with volume density $\rho$ and interatomic distances $a$, $b$ and $d$ in directions shown on Fig. 1a, we can find the coefficients for eq. (1) (see [14-16] for details): $M = \rho abd$, $K_b = \dfrac{2\mu a d}{(1-\nu)b}$, $F_d = \dfrac{\mu b^2 a}{2\pi d}$, where $\mu$ is the shear modulus and $\nu$ is the Poisson ratio. Now eq. (1) can be written in the form (the second term at the left hand side is presented in continuum limit approximation):

$$\frac{c^2 \rho a d}{2\pi} \frac{\partial^2 (2\pi u/b)}{\partial (tc/b)^2} - \frac{2\mu a d}{2\pi(1-\nu)} \frac{\partial^2 (2\pi u/b)}{\partial (x/b)^2} + \frac{\mu b^2 a}{2\pi d} \sin(\frac{2\pi u}{b}) = F - f , \qquad (2a)$$

where $c^2 = 2\mu/(\rho(1-\nu)) \equiv c_l^2 (1-2\nu)/(1-\nu)^2$; $c_l$ is the longitudinal acoustic velocity (or P wave velocity). Note that $c_s < c < c_l$, where $c_s$ is the shear wave velocity (or S wave velocity). The equivalent form is (for simplicity we will suppose that $a=b=d$):

$$\frac{\partial^2 (2\pi u/b)}{\partial (tc/b)^2} - \frac{\partial^2 (2\pi u/b)}{\partial (x/b)^2} + A^2 \sin(\frac{2\pi u}{b}) = (F-f)\frac{2\pi A^2}{\mu b^2} \tag{2b}$$

The dimensionless parameter $A$ has the expression $A = ((1-\nu)/2)^{1/2}$. In the derivation of eq. (2b), $A^2$ is essentially the ratio of the amplitude of two forces: one is the force amplitude between an atom and the substrate layer and the other is the force amplitude between neighbouring atoms at the top layer (both forces appear when the atom shifts horizontally). This ratio is of order unity. In the derivation of this equation, the system was treated macroscopically as an elastic continuum to relate the force constants in the FK model to the elastic constants of the crystal. To describe the respective coefficients for the situation where slip occurs between two external surfaces in contact, we will use eq. (2) with one significant change: we shall treat the parameter $A$ phenomenologically, using the result for a crystal as a guide. So we assume that $A$ likewise depends on the ratio of two relevant forces. The force amplitude (per unit area) experienced by an asperity due to neighbouring asperities along the slip direction is exactly the same as it was for the case of plasticity. But the force amplitude between asperity and substrate is different and depends on the normal stress $\Sigma_N$. Indeed, when $\Sigma_N = 0$ the force is zero, since there is no interaction between asperities and a substrate. On the other hand, when $\Sigma_N$ reaches the penetration hardness $\sigma_p$ the interface between the two blocks disappears and the corresponding force amplitude is essentially the same as in the case of plasticity. So we regard $A$ as a function of the ratio of $\Sigma_N$ to $\sigma_p$: $A = f(\Sigma_N/\sigma_p)$. We cannot determine the exact functional form of $A$ from our model alone, but $A$ must reduce to zero when $\Sigma_N = 0$ and reaches essentially its maximum of $((1-\nu)/2)^{1/2}$ when $\Sigma_N = \sigma_p$. The simplest choice is $A = ((1-\nu)/2)^{1/2} \Sigma_N / \sigma_p \approx \Sigma_N / \sigma_p$. Thus, the coefficient $A$ reflects how deeply the asperities from two opposing surfaces interpenetrate (see illustration on Fig. 2). Its value can be considered as the ratio between actual and nominal contact areas (to within a factor $((1-\nu)/2)^{1/2}$), since this area ratio is approximately $\Sigma_N / \sigma_p$ [17].

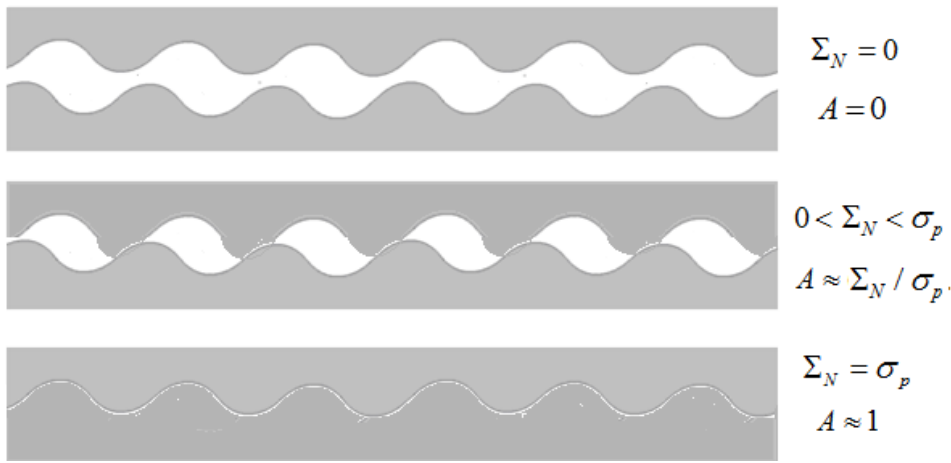

Figure 2. Schematic representation of asperity positions on frictional surfaces in cases when normal stress is 1) absent (upper panel), 2) maximal possible (bottom panel), and 3) in between (middle panel).

It is convenient to re-write eq. (2b) in dimensionless form:

$$\frac{\partial^2 u}{\partial t^2} - \frac{\partial^2 u}{\partial x^2} + \sin(u) = \Sigma_S^0 - f, \tag{3}$$

where $u$, $x$ and $t$ are now in units of $b/(2\pi)$, $b/A$ and $b/(cA)$, respectively, and $\Sigma_S^0$ and $f$ are external shear stress and friction force per unit area, both in units of $\mu A/(2\pi)$. The derivatives $\varepsilon = \partial u/\partial x$ and $w = \partial u/\partial t$ are interpreted as the dimensionless strain and the dimensionless slip in units of $A/(2\pi)$ and $cA/(2\pi)$, respectively. It is also useful to introduce the xz component of the stress tensor $\sigma_s = 2\mu\varepsilon$. Thus, the dimensionless stress is measured in units $\mu A/\pi$.

### 3. Solution of the sine-Gordon equation

In the absence of driving and dissipative forces the right-hand side of eq. (3) is zero and the latter reduces to the well-known and thoroughly-investigated SG equation. Here we consider some solutions appropriate for our problem.

*Periodic solution*

We look for nonlinear wave solutions in the form $u = u(\theta)$, where $\theta = k\xi$, $\xi = x - Ut$, $k$ is the wave number measured in units of $A/b$, and $U$ is the wave velocity in units of $c$. Integrating eq. (3) and restricting ourselves to periodic solutions with $|U|<1$ we obtain

$$u = \arcsin[\pm cn(\beta \cdot \xi)], \; \beta = [m(1-U^2)]^{1/2},$$
$$\sigma_s = 2\beta \cdot dn(\beta \cdot \xi), \; w = U\sigma_s, \; k = 2\pi N = \frac{\pi\beta}{2K}, \tag{4}$$

where $dn$ and $cn$ are the elliptic Jacobi functions, $N$ is the density of kinks in units of $A/b$, and $K(m)$ is the complete elliptic integral of the first kind of modulus $m$ ($0 \leq m \leq 1$). Solution (4) describes an infinite sequence of kinks of one sign and of constant density $N$, moving with constant velocity $U$. In what follows the terms kink and dislocation will be used interchangeably. It is useful to introduce three variables averaged over an oscillation period,

$$W = \oint \frac{w d\theta}{2\pi} = \frac{UN}{2\pi}, \; \Sigma_S = \oint \frac{\sigma_s d\theta}{2\pi} = k, \; \mathrm{E} = \oint \frac{\varepsilon d\theta}{2\pi} \tag{5}$$

These variables correspond to the measurable parameters of slip velocity, stress and strain. The parameters of a dislocation (amplitude of stress $\sigma_s^0$ and strain $\varepsilon^0$ associated with the presence of the dislocation) are: $\sigma_s^0 = \mu A/\pi$, $\varepsilon^0 = A/(2\pi)$. The characteristic width $D$ of a dislocation is also an important parameter and can be expressed by simple relation: $D \approx 2\pi b/A$. This width is usually much larger than the distance between asperities (see estimate below), which justifies the use of the continuum limit of the FK model.

A slip pulse may include one or more dislocations. In the context of friction the formulae (4, 5) may be interpreted as follows.
1. Dislocations, which are areas of accumulated stress, are nucleated on the surface by an applied shear stress in the presence of pre-existing local surface inhomogeneities. The presence of a positive dislocation is associated with a local stretch of the material above the frictional surface and a local compression of the material below the frictional surface. The presence of a negative dislocation has just the opposite effect. Thus the strain and stress anomalies associated with the presence of a dislocation are anti-symmetrical relative to the frictional surface.

2. In crystals, the displacement of a dislocation under shear stress (its "mobility") is much larger than the corresponding translation of the entire atomic plane. In the same way, the mobility of a macroscopic dislocation over the "bumpy road" on a frictional surface is much larger than the mobility of the whole surface. In both cases, the displacement of a dislocation (a pre-stressed area) requires much less external stress. So the relative sliding of two bodies along a planar interface occurs due to movement of dislocations. The passage of a dislocation through a particular point on the sliding surface shifts the contacted bodies locally by a typical distance *b*.

3. A dislocation may propagate with any velocity *U* ranging from 0 to *c*, in particular it may be stationary.

4. The average velocity of sliding, i.e. the observable slip rate *W*, is proportional to dislocation velocity *U* and dislocation density *N*.

5. The parameters of a dislocation (stress amplitude and pulse width) are entirely defined by the material parameters and normal stress and do not depend on process parameters such as dislocation density and slip rate.

*Non-stationary solution*

To model the transition from static to dynamic friction one needs a non-stationary solution of the SG equation. Since in many practical cases the actual position of any particular dislocation is not important, the dynamics of a sequence of a large number of dislocations may be described by variables averaged over an oscillation period, such as eq. (5). Witham developed a technique allowing one to construct a system of equations which includes averaged variables that are slowly varying in time and space [18]. The general solution of the Witham equations superimposed on the SG equation has been found by Gurevich et al. [19] and is described by Gershenzon [16] in a form more appropriate for our use. Let us consider the so-called "self-similar simple wave" solution [16, 20]:

$$\frac{x}{t} = V(m), \quad V = \frac{G-\alpha}{G+\alpha}, \quad U = \frac{\varsigma-\alpha}{\varsigma+\alpha}, \quad \Sigma_s = 2\pi N = \frac{\pi(\varsigma+\alpha)}{2K\sqrt{m\varsigma\alpha}}, \quad W = 2\pi UN, \qquad (6)$$

where $G = \varsigma(E - K\sqrt{m_1})/(E + K\sqrt{m_1})$, $\varsigma = (1-\sqrt{m_1})^2/m$, $m_1 = 1-m$, *E* is the complete elliptic integral of the second kind and *α* is a constant ($0 < \alpha \leq 1$) determined by the problem. The variable *V* is the nonlinear group velocity in units of *c*. Along a line *x/t*=*V*=constant in the *x-t* plane, all variables are constant. The solution is represented by a region expanding in time and bounded between the lines *x/t*=*V*(*m*=0)=$V^-$=−1 and *x/t*=*V*(*m*=1)=$V^+$. Here and below the indices + and − designate the leading and trailing edges, respectively. Note that inside the expending region all variables are functions of time and position.

Let's consider the following problem. Suppose the point *x*=0 divides the areas of stressed (*x*<0) and unstressed (*x*>0) material ($\Sigma_s(t=0, x<0) = \Sigma^-$, $\Sigma_s(t=0, x>0) = \Sigma^+ = 0$) and assume that at time *t*=0 the external shear stress reaches the value necessary to overcome static friction (note that here $x=0$ does not necessarily coincide with the edge of a sample). The transition from the static to dynamic case is described by formulas (6) (above) and (7) [16, 20]:

$$\alpha = \varsigma^-, \varsigma^- \equiv \varsigma(m^-), V^+ = \frac{1-\varsigma^-}{1+\varsigma^-}, V^- = \frac{G^- - \varsigma^-}{G^- + \varsigma^-}, G^- \equiv G(m^-), \quad (7a)$$

where $m^-$ is defined by the transcendental relation $\Sigma^- = \dfrac{\pi}{2K^-\sqrt{m^-}}, K^- \equiv K(m^-)$. It is also useful to derive a formula for the stress drop [20]:

$$\Delta\Sigma = \frac{E(m^-)}{E(m^-) + K(m^-)} \quad (7b)$$

Figure 3 schematically shows the trajectory of detachment fronts in the (x-t) plane for an initial value of the $\Sigma_S/\Sigma_N$ ratio. In this particular example both velocities are less than the shear velocity. If the $\Sigma_S/\Sigma_N$ ratio is large enough, the detachment velocity may be larger than the shear velocity and may approach the value $c$. Figure 4 illustrates the spatial distribution of process parameters for t=0 and t>0.

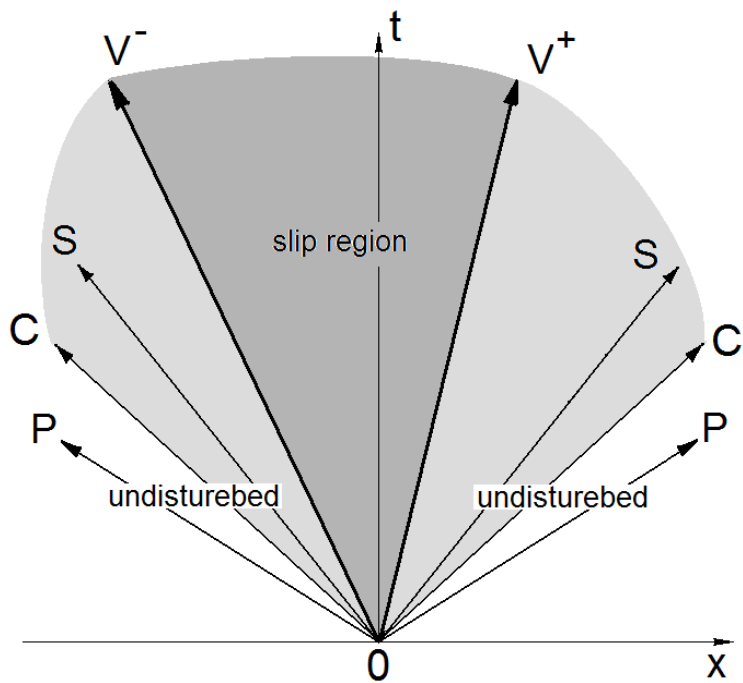

Figure 3. Trajectory of detachment fronts in the (x-t) plane, moving with velocity $V^-$ to the left and $V^+$ to the right. The pulse region is indicated by the dark gray colour. $P$ and $S$ denote the trajectories of $P$ and $S$ waves if they were emanating from the initial point. $C$ indicates the maximum velocity $c$ which the detachment fronts can have if the initial shear stress is infinite. The gray areas (both dark and light regions) denote where slip can theoretically occur.

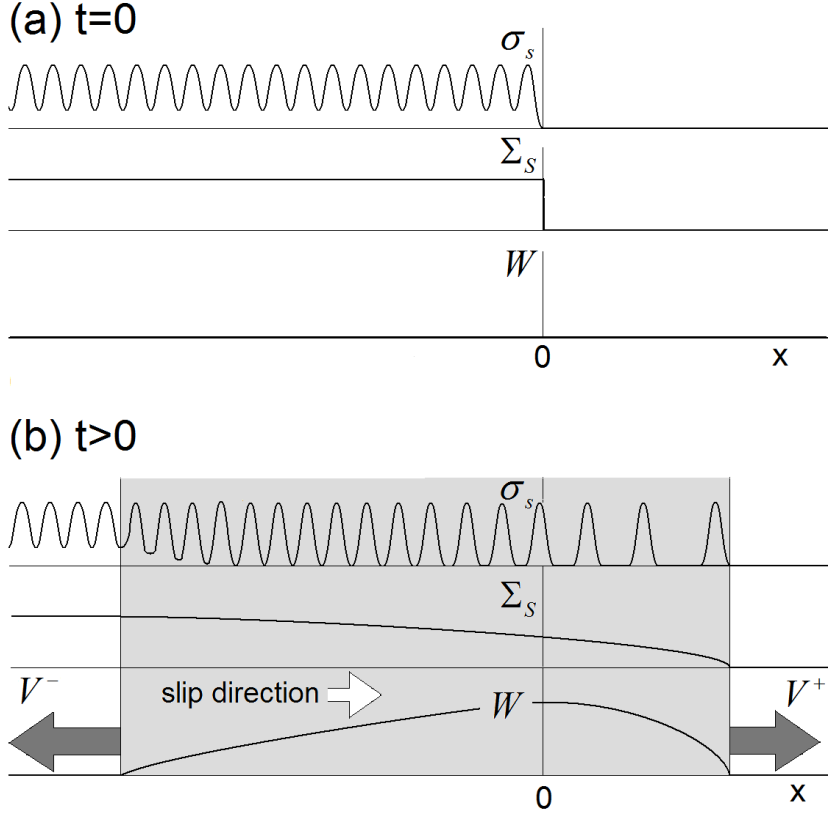

Figure 4. Schematic of spatial distribution of shear stresses $\sigma_S$ and $\Sigma_S$ as well as slip velocity $W$ (a) just before a slip pulse initiation (t=0) and (b) during slip pulse propagation (t>0). The dark arrows show the directions and relative magnitudes of the "fast" (left) and "slow" (right) fronts. The slip area (shown in light grey colour) is bounded by the fronts, while the slip direction, shown by the white arrow, is directed from the stressed to unstressed area. The maximum slip velocity, hence maximum total slip, occurs at $x = 0$.

The velocity of dislocation movement $U(x,t)$ (formulae (6) and (7)) ranges from zero at the pulse trailing edge to the value $V^+$ at the pulse leading edge. The movement of dislocations is accompanied by slip with velocity $W(x,t)$, eq. (6). The slip velocity equals zero at the trailing and leading edges of a pulse and has a maximum value at $x=0$. Assuming $V(x=0)=0$ we find from eq. (6) and (7) the maximum slip velocity:

$$W(x=0) = \frac{\pi(\varsigma^0 - \varsigma^-)}{2K\sqrt{m^0 \varsigma^0 \varsigma^-}}, \tag{8}$$

where $\varsigma^0 \equiv \varsigma(m^0)$ and $m^0$ is a solution of the equation $G(m^0) = \varsigma^-$.

## 4. Modelling of experiments

Figure 5 shows the results of a typical experiment with progressive increase of shear force with constant normal force [8]. As one can see the first slip pulse appears at a force ($F_S^{crit}$) about a one third of the value predicted by the conventional friction theory ($F_S^{theory}$).

Indeed, since $F_S^{theory} = \mu_{static} F_N \approx F_N \mu/(2\pi\sigma_p) \approx (0.4 \div 0.8) F_N$, (where $\mu_{static}$ is coefficient of static friction) then $F_S^{theory}/F_S^{crit} = (0.4 \div 0.8) F_N / F_S^{crit} \approx 2 \div 4.5$. Why is the actual critical shear force much less than the theoretical one? The obvious reason is the presence of inhomogeneities in the spatial shear stress distribution. A specific explanation in our model is the appearance of macroscopic dislocations. How and why they do appear? As long as $F_S < F_S^{crit}$, the friction force, i.e. adhesion between asperities, prevents movement of the upper (mobile) block. The tangential force acting between two asperities from the opposite surfaces is $1/A$ times larger than the force acting (over an equal area) on internal atomic plane placed inside block. Due to this "force multiplication" and randomness of the contact area, there is always the possibility that for some contacts the force exceeds the value necessary to unlock an asperity and it jumps over an opposing asperity. The result is a local stress redistribution and the appearance of a macroscopic dislocation, where n+1 or n-1 asperities on one frictional surface are placed over n asperities of the opposite surface. So dislocations may appear under an external shear stress even less than $F_S^{crit}$, i.e. before the first observable sliding pulse. Due to the action of this force the upper block (Fig. 1a) is elastically deformed, primarily from the trailing (rear) edge, by an amount $\Delta x$. Deformation of the upper block just before the first slip pulse can be estimated by the relation: $\Delta u = \Delta x F_S^{crit}/(2\mu \Delta x \Delta y)$, where $\Delta y$ is the size of the upper block in the y direction and $\Delta x \cdot \Delta y$ is the area of the frictional surface which "absorbs" this deformation. So the maximum number of dislocations which may formed before the first pulse is $\Delta u / b = F_S^{crit}/(2\mu \Delta y b) \approx 30$. Here we used the following values: $b = 1\mu m$, $F_S^{crit} = 0.6 kN$ (see Fig. 5) and $\mu = 1.7 GPa$. Physically, dislocations appear under applied stress because they can move much more easily then a rigid surface. The passage of a dislocation through a particular point shifts the frictional surfaces relative to each other, thus providing a local sliding.

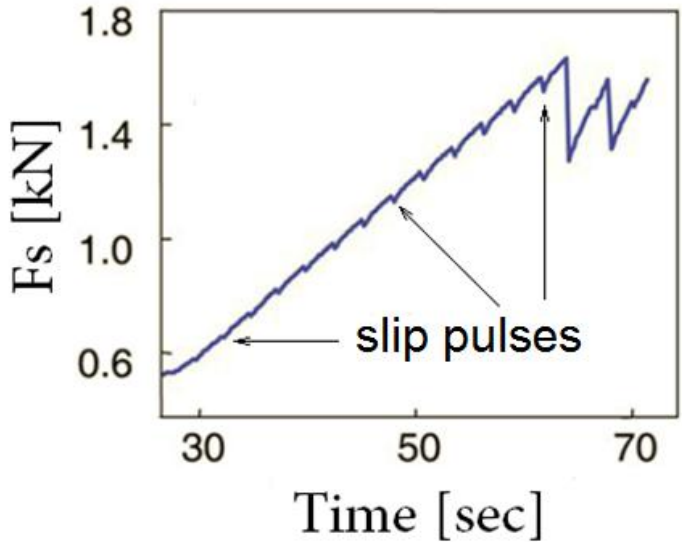

Figure 5. The main event triggering sliding is preceded by a sequence of frustrated sliding pulses (crack-like precursors). The curve shows $F_S$ as a function of time for a 200 mm long sample loaded at $F_N = 3.3$ kN. (from [8])

Now let's return to eq. (3). In the static case, i.e. before the first pulse and between pulses, the right-hand side of eq. (3) should be zero, i.e. static friction force per unit area is equal to shear stress ($f \equiv f_{static} = \Sigma_S^0$). If this were not the case, then dislocations would move with increasing velocity, which is not true since slip is not observed between pulses. So the periodic solution (formulae (5) and (6)) ) with *U=0* should completely describe both microscopic and observable parameters, such as size and density of dislocations and internal stress. Supposing that $\sigma_p = 400$ MPa (for PMMA material) and $\Sigma_N$ ranges from 2.5 to 10 MPa (as in experiments [7-10]), we find that *A* ranges from $0.4 \cdot 10^{-2}$ to $1.5 \cdot 10^{-2}$. Using the range of *A* values we find the range of dislocation widths to be *D* = 0.4 - 1.7 mm, i.e. the width of a dislocation is from 250 to 1000 times the distance between asperities.

Formulae (5) and (6) can also be used to describe the uniform sliding of two blocks. In this case $f \equiv f_{dynamic} = \Sigma_S^0$ but *U* is not zero. The latter can be found if the sliding velocity and stress state are known.

However, the most interesting part of these experiments lies in the dynamics of a slip pulse. A pulse occurs when the dislocation density , hence the internal stress, reaches the critical value and the friction force is not able to suppress dislocation movement. As has been shown experimentally, a pulse is initiated by the detachment front [7]. From the view point of our model it means that the first dislocation starts to move toward the leading edge, tearing down the adhesion contacts. The elastic energy accumulated inside the blocks, hence the repelling force between dislocations, drives the movement of the latter. The static friction became dynamic friction. The latter occurs due to a variety of dissipation mechanisms including plastic deformation, wear, and excitation of phonon modes (heat) [17]. Dynamic friction is usually less than static friction but of the same order of magnitude. In most cases the elastic energy accumulated and released during frictional cycling is larger than the energy dissipated by dynamic friction [21]. Based on the last two remarks we can expect that the right-hand side of eq. (3) remains small even for non-uniform slip, and may be ignored in the first approximation. Thus, we can use the non-stationary solutions of the SG equation obtained above to describe the dynamics of elastic energy redistribution during the transition process, i.e. the dynamics of a slip pulse. Indeed, the quantity $V^+$ is the velocity of the leading edge and the velocity of the first dislocation. This is the velocity of the detachment front propagating from the stressed to the unstressed area. The velocity *V* also may be interpreted as a detachment front velocity, however in contrast to $V^+$, it is the velocity of the disturbance propagated through the stressed area. These velocities are uniquely defined by the ratio of shear to normal stress (see curves in Fig. 6); however their values are very different and *V* is always larger than $V^+$. These two types of detachment fronts may explain the observed sudden changes of front velocity [7]. Indeed, when the detachment front propagates through the stressed area the velocity is *V*, and when it reaches the unstressed area the velocity becomes $V^+$. Figure 6 also includes the experimentally measured velocities as a function of measured ratio of shear to normal stress [10]. One can see that the experimental values are centred around the *V* curve. It is interesting that even though there is no non-zero lower limit (in our model) of the ratio of shear to normal stress which may activate transition, an effective lowest limit of about 1/3 may be seen in Fig. 6.

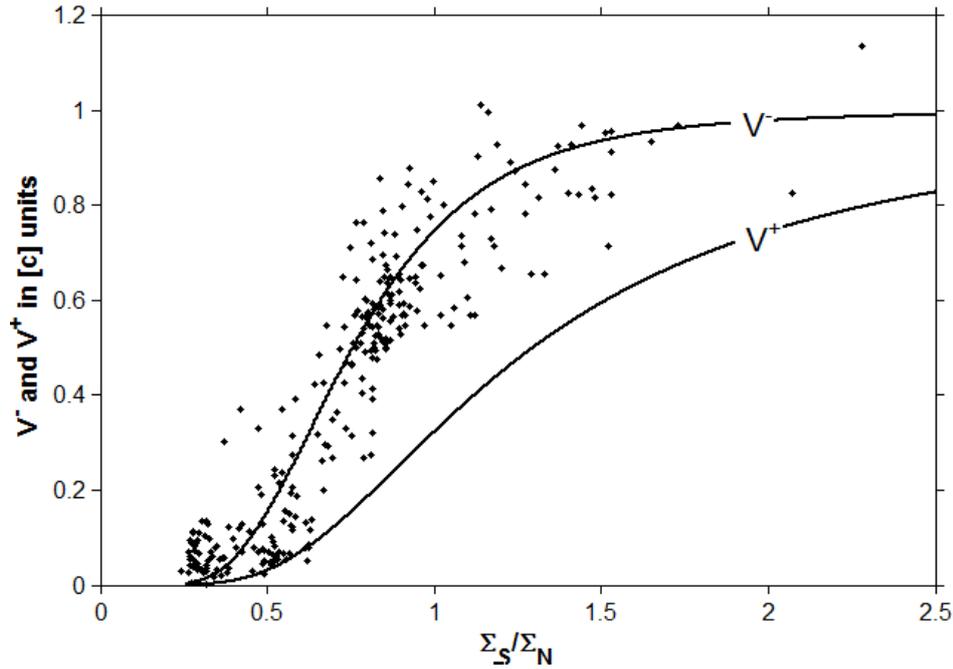

Figure 6. Detachment front velocities as a function of the ratio of shear to normal stress. Curves show theoretical results and dots experimental results (from the article [10]).

Immediately after the detachment fronts passes through a particular area, dislocations are free to move (or rather diffuse) from the stressed to the unstressed area, redistributing the accumulated shear stress. When the shear stress, i.e. dislocation density, falls to the value $\Sigma^- = \mu_{dynamic}\Sigma_N$, (where $\mu_{dynamic}$ is the coefficient of dynamic friction) slip ceases. The velocity of dislocation movement $U(x,t)$ (formulae (6) and (7)) ranges from zero at the pulse trailing edge to the value $V^+$ at the pulse leading edge. The movement of dislocations is accompanied by slip with velocity $W(x,t)$, eq. (6), orders of magnitude less than the velocity of a detachment front. Figure 7 depicts slip velocity as a function of detachment front velocity (Fig. 7a) and shear stress (Fig. 7b). Note that the same value of detachment front velocity may be accompanied by different slip velocities (and vice versa), depending on the value of $A$ (hence normal stress). However the shear stress uniquely defines slip velocity regardless of the value of normal stress (see Fig. 7b). In many cases (such as earthquakes) the shear-stress drop ($\Delta\Sigma$) after a slip pulse is an important parameter. In the framework of our model, stress drop is connected to kinematic parameters. Figure 8 shows the relation between the relative and absolute values of stress drop as a function of the detachment front velocity (see formula (7b)). It is Interesting that the stress drop does not exceed 50% of the accumulated shear stress (see Fig. 8a)

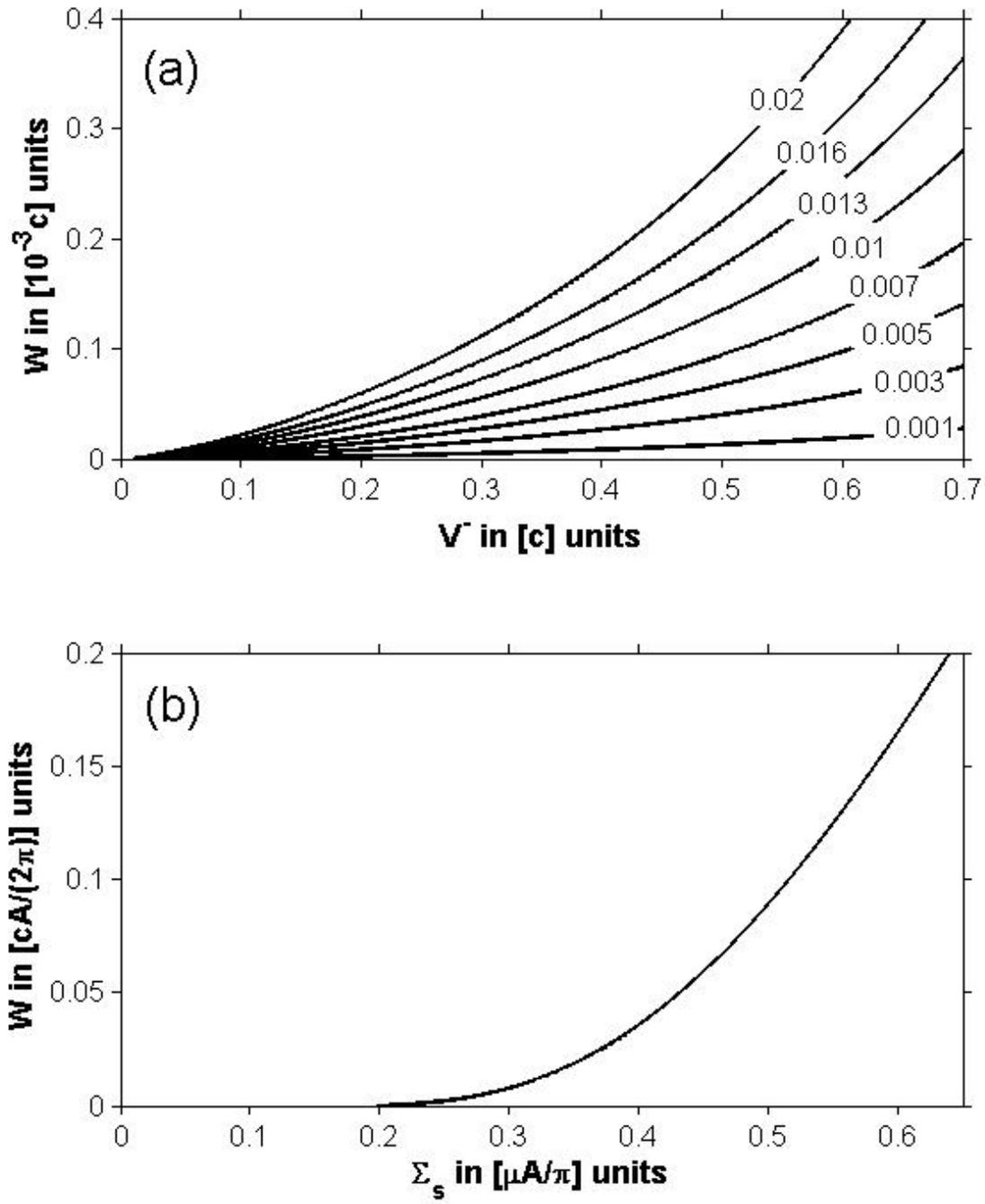

Figure 7. Slip velocity as function of (a) "fast" detachment front velocity and (b) shear stress.

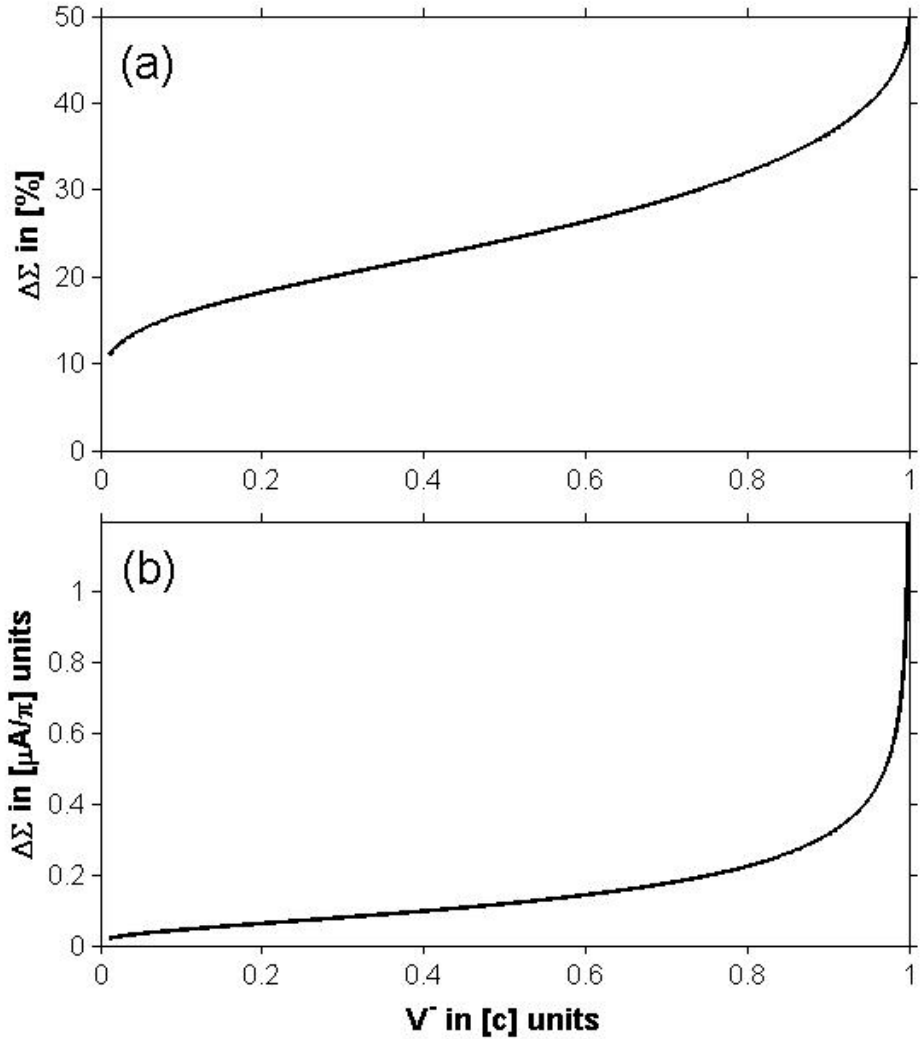

Figure 8. (a) Percent stress drop and (b) absolute stress drop, as a function of the "fast" detachment front velocity.

## 5. Discussion and conclusion

In the framework of the model developed here, a rupture-like slip occurs due to the appearance and motion of a macroscopic dislocation or train of dislocations. This approach is consistent with the "pulse-like" and "train-of-pulses" modes of dynamic rupture observed during earthquakes and described theoretically, e.g. [22] and references therein. The FK model describes quantitatively the dynamics of the transition process from static to dynamic friction in the case of slip (not of sliding, when slip edges coincide with sample edges). Simple transcendental algebraic relations between rupture velocities, slip velocity, shear-stress drop, and shear and normal stress are obtained (formulae (4)-(8) and Fig. 6 - 8). The significant consequences of this model are (1) the velocity of the detachment front (rupture) depends only on the ratio of shear to normal stress; (2) the velocity of slip depends only on shear stress and does not depend on normal stress; and (3) neither velocity depends explicitly on the friction coefficient, although the initiation and termination of the slip are defined by the static and dynamic friction coefficients, respectively. Note that the model cannot address the value of the critical threshold for the transition. Where comparisons can be made, model predictions are in good agreement with experiments (see Fig. 6). The model predicts the relations between slip velocity and detachment front velocity (Fig. 7a) and between slip velocity and shear stress (Fig. 7b). These predictions could be examined experimentally.

A possible geophysical application of the model is the description of the dynamics of transform and subduction faults, i.e. regular and slow earthquakes. A slip pulse may contain a single dislocation, as in the phenomenon of Episodic Tremor and Slip observed in Cascadia and Nankai subduction faults [23], or it may be a sequence of closely placed pulses, as in large crustal earthquakes [20]. Note that in the latter case and in the case of stick-slip experiments, dislocations are almost non-resolvable from each other. Figure 4 mimics the rupture process during an earthquake in transform faults, and figures 6, 7 and 8 quantitatively describe the relations between kinematic and dynamic parameters. Let us illustrate the applicability of the model considering, as an example, the 2004 M=6 Parkfield earthquake. The reconstruction of the earthquake kinematic parameters using various data and methods reveals that the rupture velocity ($V$) was about 3.0 km/s and the slip velocity in the strike direction at the hypocenter area ($W$) was about 0.5 m/s (see [24] and references therein). Supposing that the $P$-wave velocity was $c_l = 6$ km/s and that $v = 0.3$, we determine the value of the parameter $c$ was 5.4 km/s (see definition of $c$ below eq. (2a)). Now we can find the value of $A$ to be 0.007 from Fig. 7a. So the ratio of normal stress to penetration hardness is about 0.007. From this we can find 1) the absolute magnitude of the shear stress initiating the earthquake (also known as the yield stress [24]) and 2) the stress drop, using respectively Fig. 7b and 8b. Supposing $\mu = 30$ GPa we find $\Sigma_S = 36$ MPa and $\Delta\Sigma = 10$ MPa. Finally, we can estimate the absolute magnitude of the normal stress using Fig. 6, giving us the value of about 45 MPa for this particular case. Thus, using kinematic parameters we are able to estimate dynamic parameters. For comparison, sophisticated dynamic modelling of the Parkfield earthquake based on near-source ground motion data and using multiple approaches [24] results in a yield stress of about 31 MPa, a stress drop of 10 MPa and a normal stress of about 60 MPa, which are close to our estimates.

## Acknowledgement

We are thanks J. Fineberg for helpful discussions and comments and O. Ben-David for the experimental data. This work was supported by NSF grant #1113578.